# THE INFLUENCE OF DIVERSITY ON THE MEASUREMENT OF FUNCTIONAL IMPAIRMENT: AN INTERNATIONAL VALIDATION OF THE AMSTERDAM IADL QUESTIONNAIRE IN 8 COUNTRIES

Mark A. Dubbelman [a*], Merike Verrijp [a], David Facal [b], Gonzalo Sánchez-Benavides [c], Laura J.E. Brown [d], Wiesje M. van der Flier [a,e], Hanna Jokinen [f], Athene Lee [g], Iracema Leroi [d], Cristina Lojo-Seoane [b], Vuk Milošević [h], José Luís Molinuevo [c], Arturo X. Pereiro Rozas [b], Craig Ritchie [i], Stephen Salloway [g], Gemma Stringer [d], Stelios Zygouris [j,k], Bruno Dubois [l], Stéphane Epelbaum [l], Philip Scheltens [a], Sietske A. M. Sikkes [a,e]

**Affiliations**

[a] Alzheimer Center Amsterdam, Department of Neurology, Amsterdam Neuroscience, Vrije Universiteit Amsterdam, Amsterdam UMC, Amsterdam, The Netherlands

[b] Department of Developmental Psychology, University of Santiago de Compostela, A Coruña, Spain

[c] BarcelonaβBeta Brain Research Center (BBRC), Pasqual Maragall Foundation, Barcelona, Spain

[d] Faculty of Biology, Medicine and Health, University of Manchester, Manchester Academic Science Centre, Manchester, United Kingdom

[e] Department of Epidemiology and Biostatistics, Amsterdam UMC, Amsterdam, The Netherlands

[f] Clinical Neurosciences, Neurology, University of Helsinki and Helsinki University Hospital, Helsinki, Finland

[g] Butler Hospital, Warren Alpert Medical School of Brown University, Providence, RI, United States of America

[h] Clinic of Neurology, Clinical Center Niš, Niš, Serbia

[i] University of Edinburgh, Edinburgh, United Kingdom

[j] School of Medicine, Aristotle University of Thessaloniki, Thessaloniki, Greece

[k] Network Aging Research, Heidelberg University, Heidelberg, Germany

[l] Département de neurologie, Institut de la Mémoire et de la Maladie d'Alzheimer (IM2A) of the Pitié-Salpêtrière Hospital & ARAMIS, Sorbonne Université, Inria de Paris, Institut du cerveau et de la moelle épinière (ICM), Paris, France

**Corresponding author**

M. A. Dubbelman, MSc
Amsterdam UMC
Alzheimer Center, Department of Neurology
P.O. Box 7057
1007 MB Amsterdam
The Netherlands
Phone: +31 20 4448527
E-mail: m.dubbelman@amsterdamumc.nl

**Keywords:** instrumental activities of daily living; dementia; Alzheimer's disease; functional decline; cross-cultural validation; diversity; item response theory; differential item functioning

**Abstract:** 150 words

**Manuscript:** 3474 words, excl. tables and figures

**References:** 51

**Tables & figures:** 5

**Keywords:** 8

**Research in context:** 148 words

## Abstract (150/150 words)

**INTRODUCTION:** To understand the potential influence of diversity on the measurement of functional impairment in dementia, we aimed to investigate possible bias caused by age, gender, education, and cultural differences.

**METHODS:** 3,571 individuals (67.1 ± 9.5 years old, 44.7% female) from the Netherlands, Spain, France, United States, United Kingdom, Greece, Serbia and Finland were included. Functional impairment was measured using the Amsterdam IADL Questionnaire. Item bias was assessed using differential item functioning (DIF) analysis.

**RESULTS:** There were some differences in activity endorsement. A few items showed statistically significant DIF. However, there was no evidence of meaningful item bias: effect sizes were low ($\Delta R^2$ range 0–0.03). Impact on total scores was minimal.

**DISCUSSION:** The results imply a limited bias for age, gender, education and culture in the measurement of functional impairment. This study provides an important step in recognizing the potential influence of diversity on primary outcomes in dementia research.

## 1. Introduction

Impairment in cognitively complex 'instrumental activities of daily living' (IADL), such as doing grocery shopping, managing personal finances and using mobile devices, may be one of the first symptoms of dementia [1-3]. IADL performance is related to quality of life, caregiver burden and resource utilization [4]. Moreover, IADL impairment in preclinical stages might be a predictor of progression to dementia [5, 6]. Therefore, functional impairment is an important and highly relevant outcome measure for clinical practice and clinical trials. In recently drafted industry guidelines, the U.S. Food and Drug Administration recommended the use of functional impairment as a measure for effectiveness of treatment and of disease progression [7]. It is a potential global outcome measure in dementia research [8, 9].

Because everyday functioning relates to daily life, IADL may be especially sensitive to bias caused by various factors, such as age, gender, and cultural differences. Previous studies have shown gender effects on traditional IADL instruments [9-12], as they predominantly include household activities, which may be performed more often by women. Scientific literature concerning cultural and ethnoracial diversity in the context of dementia is scarce [13, 14]. The selection of activities to include in an IADL instrument may be culture-specific. For example, in the United States it is customary to write checks, whereas in the Netherlands, people often use online banking. Mere translation of an instrument does not always account for national (cross-cultural) disparities [15, 16], and while many functional instruments have been translated into numerous languages, there is no gold standard for cross-cultural adaptation of questionnaires [17]. This emphasizes the importance of investigating potential sources of bias and their influence on item and scale level.

We aimed to study potential influences of diversity on the measurement of functional impairment using the Amsterdam IADL Questionnaire (A-IADL-Q). Specifically, we investigated item bias caused by various factors: cross-cultural differences (operationalized by using country of residence), age, gender, and education. We obtained data from eight Western countries: The Netherlands, Spain, France, United States, United Kingdom, Greece, Serbia, and Finland.

## 2. Methods

The present study included data from 3,571 individuals with a completed A-IADL-Q from memory clinics and cognition studies from eight countries: the Netherlands (Amsterdam Dementia Cohort [18] and European Prevention of Alzheimer's Dementia Longitudinal Cohort Study, EPAD [19, 20]), Spain (Compostela Aging Study [21, 22]; EPAD; and ALFA project [23]), France (INSIGHT pre-AD [24]; EPAD; and Socrates study), United States (Butler Alzheimer's Prevention Registry [25]), United Kingdom (EPAD and SAMS project [26]), Greece (Greek Association for Alzheimer's Disease and Related Disorders), Serbia (Niš Clinic of Neurology [27]), and Finland (Helsinki Small Vessel Disease study).

Participants had some degree of cognitive complaints, or had an increased genetic or neurovascular risk for cognitive decline. Participants were recruited from memory clinics, through advertisement, or from existing databanks. Inclusion criteria ranged from being cognitively normal to having a dementia-related diagnosis. Other relevant inclusion and exclusion criteria for each cohort in this study can be found in Table 1. Participants provided written informed consent, and the studies were approved by their institutional review boards, which included, in each, consent for data sharing.

Table 1

*Information about participants, in- and exclusion criteria, and information about the A-IADL-Q administration per included sample*

| **Study name** | **Amsterdam Dementia Cohort [18]** | **Compostela Aging Study [21, 22]** | **European Prevention of Alzheimer's Dementia Longitudinal Cohort Study (EPAD) [19, 20]** | **ALFA+ Study [23]** | **INSIGHT preAD [24]** | **Butler Alzheimer's Prevention Registry [25]** | **SOCRATES** | **Greek Association of Alzheimer's Disease and Related Disorders** | **Niš Clinic of Neurology [27]** | **Helsinki Small Vessel Disease study** | **SAMS Project [26]** |
|---|---|---|---|---|---|---|---|---|---|---|---|
| **Country** | Netherlands | Spain | Spain (*n* = 218)<br>France (*n* = 103)<br>Netherlands (*n* = 88)<br>United Kingdom (*n* = 71) | Spain | France | United States of America | France | Greece | Serbia | Finland | United Kingdom |
| **Participants included** | 1,429 | 600 | 480 | 333 | 308 | 154 | 98 | 61 | 45 | 43 | 22 |
| **Age range** | 25–84 years | 50–101 years | 51–88 years | 49–73 years | 70–85 years | 58–77 years | 46–85 years | 65–92 years | 26–93 years | 66–75 years | 65–82 years |
| **Research environment** | | | | | | | | | | | |
| *Recruitment* | Consecutive memory clinic patients | MCI patients referred by GP | Participants from existing study cohorts | Mostly offspring of AD patients | Consecutive memory clinic patients & advertisement recruited | Advertisement recruited | Memory clinic patients | Patients from day center for dementia | Memory clinic patients | Patients with neuroimaging data selected from existing databank | Recruited from dementia research registry, memory clinic patients |
| *Relevant inclusion and exclusion criteria* | None | Cognitive complaints without dementia;<br>Age ≥ 50 years | No dementia;<br>Age ≥ 50 years | CN (MMSE ≥ 26, CDR 0);<br>No neurological diseases;<br>Age 45–74 years | CN (MMSE ≥ 27, CDR 0);<br>Amyloid PET at baseline;<br>No episodic memory deficits, no neurological diseases;<br>Not living in nursing home;<br>Age 70–85 years | CN or mild memory loss;<br>No neurological diseases or dementia diagnosis;<br>Age 55–85 years | Dementia-related diagnosis (MMSE ≥ 10);<br>No neurological diseases other than dementia;<br>Age 40–85 years | Dementia-related diagnosis;<br>Reliable informant;<br>No neurological diseases other than dementia;<br>Age ≥ 65 years | CN, MCI, post-stroke cognitive impairment | No major neurological symptoms or psychiatric disease;<br>Independence in basic ADL;<br>No large infarcts, hemorrhages, contusion or tumor on MRI;<br>Age 65–75 years | SCD (ECog ≥ 1.436 and answered "yes" when asked if "concerned they have a memory or other thinking problem"), MCI;<br>Age ≥ 65 years |
| **A-IADL-Q version** | Original (*n* = 730)<br>SV (*n* = 699) | Original | Original | SV | Original | SV | SV | Original (*n* = 28)<br>SV (*n* = 33) | SV | SV | Version adapted from original |
| **Clinical measures** | | | | | | | | | | | |
| *Participants (%) selected for validation** | 1,369 (95.8) | 300 (50.0) | 480 (100.0) | 333 (100.0) | 308 (100.0) | 154 (100.0) | — | 26 (42.6) | 45 (100.0) | 43 (100.0) | 22 (100.0) |
| *Measures available* | MMSE, CAMCOG, CDR, GDS | MMSE, CAMCOG, GDS | MMSE, CDR, GDS | MMSE, CDR | MMSE, CDR | MMSE, GDS | None | MMSE, CAMCOG, CDR, GDS | MMSE | MMSE, GDS | GDS |

* Participants living in a nursing home were excluded from validation and no clinical measures were obtained for them.

Additional inclusion and exclusion criteria are available upon request.

*Abbreviations:* MCI mild cognitive impairment; GP general practitioner; AD Alzheimer's disease; CN cognitively normal; MMSE Mini-Mental State Examination; CDR Clinical Dementia Rating; ADL activities of daily living; MRI magnetic resonance imaging; SCD subjective cognitive decline; CAMCOG Cambridge Cognitive Examinations; GDS Geriatric Depression Scale

## 2.1 Measures

### 2.1.1 Amsterdam IADL Questionnaire (A-IADL-Q)

The Amsterdam IADL Questionnaire (A-IADL-Q) assesses cognitively complex IADL that are prone to decline in incipient dementia. It covers a wide range of activities: the original version contains 70 items, while the short version (A-IADL-Q-SV) has 30. Both the original and short version were used in the included studies. We analyzed both versions, with a special focus on the short version, because all items from the short version are also included in the original, and can therefore be compared between all participants.

Unlike many other IADL instruments [28], the A-IADL-Q has been extensively validated and has been shown to have good internal consistency, validity and reliability [29-31]. Furthermore, it appears to be independent of age and gender [30], and sensitive to change over time [32]. The short version was developed to create a more concise measure, as well as to reduce potential cultural bias by only including widely relevant activities [33]. International use of the A-IADL-Q is steadily increasing. All translations have gone through a cross-cultural adaptation process based on procedures described by Beaton, Bombardier [34], in which experts and prospective users were asked to evaluate the translated instrument (a more detailed description of this process can be found in the Supplementary Material).

The questionnaire is scored using item response theory (IRT), as described elsewhere [29, 31]. IRT assumes that an instrument measures a latent trait, which is represented in a scale ranging from total absence to abundance of the particular trait [35]. The A-IADL-Q latent trait is 'IADL functioning' [30]. In IRT, parameters are calculated for each item, which contain information about item response category location (or difficulty, i.e., at which trait level half the population endorses a given response category of an item), as well as slope (or discriminatory ability, i.e., how well an item can distinguish between people with lower and higher levels of the trait).

All A-IADL-Q items have five response categories, ranging from having 'no difficulty' in performing an activity to being 'unable to perform' an activity due to cognitive problems. IRT-based *T*-scores representing the trait level were calibrated in a memory-clinic population and were centered around a mean of 50 with a standard deviation (SD) of 10. Lower scores indicate more severe functional impairments.

#### 2.1.2. Clinical Measures

Mini-Mental State Examination (MMSE, scores range 0–30) [36] and Cambridge Cognition Examination (CAMCOG, scores range 0–107) [37] served as general indications of cognitive functioning. For both measures, lower scores indicate worse cognition. The Clinical Dementia Rating (CDR) [38] was an indicator of functional status. A global CDR score of 0 represents no dementia, and scores of 0.5 to 3 are related to more advanced stages of dementia (and thus more functional impairment). Lastly, the short form Geriatric Depression Scale (GDS, scores range 0–15) [39] was used to assess depressive symptoms, where higher scores are indicative of more severe depressive symptoms. Data was not obtained for all included participants: we excluded individuals living in nursing homes ($n$ = 130) because they have limited IADL independence.

### 2.2 Statistical Analyses

We investigated item bias using 'differential item functioning' (DIF) analysis. DIF analysis is a technique for identifying items that have different item locations and/or slopes in different groups. DIF is assumed to occur when the relationship between a test item and the latent trait is not the same across study-irrelevant groups [35]. It is considered a variation in measurement and is therefore undesirable [40]. We studied DIF in the following groups: (1) nationality, using the Dutch cohort as a reference group, while grouping all other studies by country; (2) men and women; and, based on median split, (3) young (< 67.2 years) and old age (≥ 67.2 years); and (4) low (< 12 years) and high education (≥ 12 years).

For all DIF analyses, a minimum count of one case in at least two different response categories was required in each group for every item. We used the ordinal logistic regression (OLR) approach, which is often used and can be performed in standard software. OLR has previously been shown to be superior the Mantel Haenszel procedure [41]. We used the 'lordif' package version 0.3-3 for R, developed by Choi, Gibbons [40]. 'lordif' has been used extensively in the literature, assuring appropriateness and replicability of our procedures. In the OLR approach, a null model and three hierarchically nested models are created and compared for each item. When DIF is present and constant across all levels of the latent trait, it is called uniform DIF. The response categories of an item with uniform DIF are located at a different location in each group [42]. When an item is easier at one level of the trait and more difficult at another level, it is considered to have non-uniform DIF [42]. Items with non-uniform DIF have different discriminatory abilities in each group. Statistically significant DIF was determined on the basis of the likelihood-ratio chi-square test with an $\alpha$ level of .01, to avoid type I error, and because multiple nested models are being tested for each item. Because of inflated type I error in OLR DIF analyses [43], we added a step to establish presence of practically meaningful DIF [44, 45], based on a McFadden's pseudo $R^2$ ($\Delta R^2$) value of .035 or larger. This approach reduces the risk of finding significant but negligible DIF, albeit at the cost of a reduction in power [43]. Furthermore, we used the following effect size criteria to quantify DIF size: $\Delta R^2$ values between .035 and .070 for moderate, and above .070 for large DIF [43]. To refine DIF detection and effect size estimates, we then performed Monte Carlo simulations over 1,000 replications in which the detection criteria as well as effect size measures are computed repeatedly over simulated data based on the empirical datasets. The simulated data are generated under the hypothesis that there is no DIF, while keeping the observed group differences in trait levels.

As a means of construct validation, Pearson's $r$ for continuous or Kendall's $\tau$ correlation coefficients for ordinal-level measures were calculated for the association between A-IADL-Q-SV $T$-scores and age, education level, gender of the participant, cognitive functioning (MMSE and CAMCOG), functional state (CDR), and mood (GDS).

Data were processed in SPSS Statistics version 22 [46] and R version 3.6.1 [47].

## 3. Results

On average, participants were 67.1 ± 9.5 (*m* ± SD) years old. Table 2 shows the demographics and clinical measures of all participants, as well as stratified by country.

Table 2. Demographics and clinical characteristics for all participants, and grouped per country.

| | **All** | **Nether-lands** | **Spain** | **France** | **United States** | **United Kingdom** | **Greece** | **Serbia** | **Finland** |
|---|---|---|---|---|---|---|---|---|---|
| **Total *n*** | 3,571 | 1,515 | 1,151 | 509 | 154 | 93 | 61 | 45 | 43 |
| **Females, n (%)**[1] | 1,597 (44.7) | 637 (42.0) | 485 (42.1) | 262 (51.5) | 104 (67.5) | 43 (46.2) | 18 (29.5) | 25 (55.6) | 23 (53.5) |
| **Age, years** | 67.14 ± 9.5 | 63.78 ± 8.5 | 67.84 ± 10.4 | 73.48 ± 6.2 | 66.65 ± 4.5 | 68.42 ± 5.8 | 79.99 ± 6.4 | 65.44 ± 13.1 | 71.69 ± 2.8 |
| **Education years** | 12.19 ± 3.9 | 11.34 ± 3.2 | 11.97 ± 4.4 | 13.95 ± 3.7 | 16.82 ± 2.3 | 12.99 ± 3.1 | 9.50 ± 4.3 | 13.93 ± 4.3 | 12.93 ± 5.5 |
| **Dementia diagnosis, n (%)**[1] | 860 (29.9) | 647 (47.2) | 188 (20.2) | 0 (0) | 0 (0) | 0 (0) | 21 (80.8) | 4 (8.9) | 0 (0) |
| **A-IADL-Q** | | | | | | | | | |
| T-*score*[2] | 58.40 ± 14.2 | 51.54 ± 11.7 | 61.82 ± 15.2 | 67.33 ± 9.4 | 67.48 ± 3.5 | 71.16 ± 5.1 | 39.48 ± 13.9 | 61.67 ± 8.8 | 66.30 ± 5.2 |
| **Clinical measures**[1] | | | | | | | | | |
| *MMSE* | 26.20 ± 4.6 | 24.22 ± 5.0 | 27.76 ± 3.7 | 28.62 ± 1.2 | 29.35 ± 1.0 | 28.46 ± 1.5 | 19.58 ± 4.6 | 27.49 ± 3.6 | 27.60 ± 2.2 |
| *CAMCOG* | 78.57 ± 17.3 | 78.75 ± 16.1 | 80.98 ± 19.1 | — | — | — | 41.62 ± 9.7 | — | — |
| *CDR, M (IQR)* | 0 (0–0.5) | 0.5 (0–1) | 0 (0–0) | 0 (0–0) | — | 0 (0–0) | 2 (0.5–2) | — | — |
| *GDS* | 3.66 ± 3.6 | 3.80 ± 3.3 | 4.09 ± 4.0 | 4.33 ± 4.2 | 0.85 ± 1.3 | 3.52 ± 4.5 | 2.38 ± 3.1 | — | 2.10 ± 3.1 |

All data are displayed as mean ± standard deviation, except as stated otherwise. "—" denotes that data were not available. [1] Data was not obtained for all participants. [2] The score shown is based on either the original or short version of the A-IADL-Q, as administered to each participant.

*Abbreviations:* M median; IQR interquartile range; A-IADL-Q Amsterdam Instrumental Activities of Daily Living Questionnaire; MMSE Mini-Mental State Examination; CAMCOG Cambridge Cognitive Examinations; CDR Clinical Dementia Rating; GDS Geriatric Depression Scale

The overall mean score on the A-IADL-Q was 58.40 ± 14.2. A-IADL-Q scores per country are shown in Table 2.

## 3.1 Item endorsement

Table 3. Differences in endorsement in selected activities.

| | **Country** | | | | | | | | **Age** | | **Gender** | | **Education** | |
|---|---|---|---|---|---|---|---|---|---|---|---|---|---|---|
| **Activity** | **Nether-lands** | **Spain** | **France** | **United States** | **United King-dom** | **Greece** | **Serbia** | **Finland** | **Young** | **Old** | **Men** | **Women** | **Low** | **High** |
| Minor repairs | 46.2% | 57.9% | 67.4% | 55.8% | 62.4% | 55.7% | 57.8% | 72.1% | 53.6% | 55.8% | **70.9%** | **42.4%** | 52.8% | 58.8% |
| Washing machine | 58.4% | 70.7% | 77.0% | 92.2% | 75.3% | 63.9% | 71.1% | 81.4% | 72.4% | 63.5% | **44.0%** | **93.1%** | 65.7% | 73.1% |
| Withdrawing cash from ATM | 69.6% | 64.7% | 82.7% | 74.0% | 80.6% | 9.8% | 55.6% | 72.1% | 77.2% | 62.6% | 75.8% | 75.4% | **66.2%** | **82.0%** |
| Working | 52.3% | 42.4% | 54.2% | 66.9% | 24.7% | 9.8% | 53.3% | 58.1% | **61.6%** | **36.2%** | 53.1% | 50.9% | 47.4% | 55.4% |
| Using a computer | **82.0%** | **52.7%** | **75.2%** | **97.4%** | **94.6%** | **8.2%** | **53.3%** | **81.4%** | 82.1% | 60.3% | 79.7% | 73.3% | 65.3% | 84.3% |
| Public transportation | **49.7%** | **70.5%** | **86.2%** | **27.9%** | **59.1%** | **83.6%** | **51.1%** | **79.1%** | 58.0% | 64.5% | 59.0% | 66.5% | 55.4% | 68.0% |

Differences of interest between groups within each factor (country, age, gender, and education) are displayed in bold. Endorsement of other activities included in the Amsterdam IADL Questionnaire did not differ as much and these activities are not displayed here.

Generally, item endorsement was comparable between countries, as well as between men and women, younger and older participants, and participants with lower and higher education. Table 3 highlights a few activities in which there were apparent differences. ‘Minor repairs’ was endorsed by a larger percentage of men, as compared to women. Conversely, ‘using a washing machine’ was endorsed more often by women. Participants with a lower indication endorsed ‘withdrawing cash from an ATM’ somewhat less often than participants with a higher education. Older participants were less likely to work, compared to younger participants. Participants from Greece, Spain, and Serbia used computers less often than those from the other countries. Participants from the United States appeared to use public transportation less often than those from European countries (see Table 3).

## 3.2 Item bias

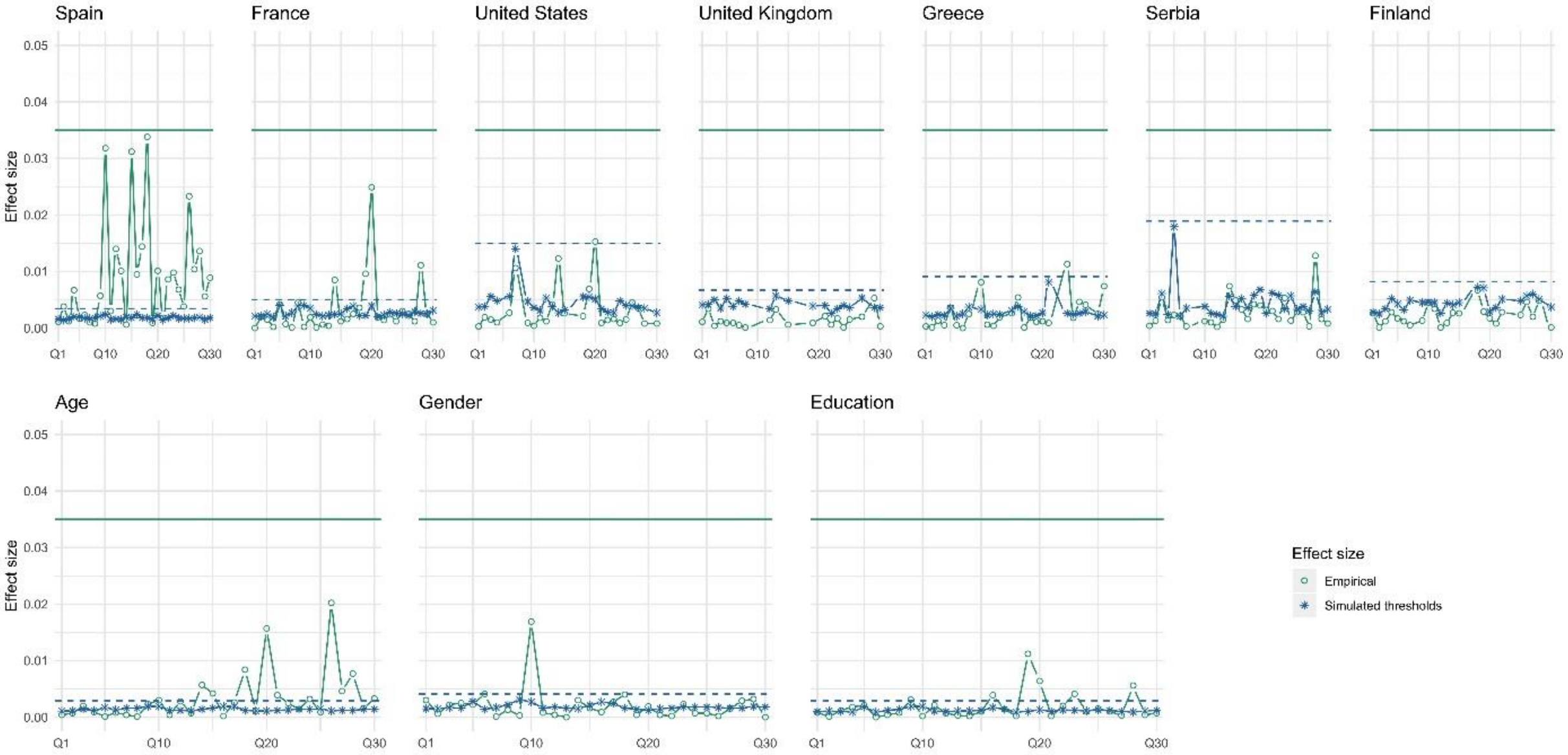

**Figure 1:** DIF effect sizes for country, age, gender, and education in the A-IADL-Q-SV.
Green circles represent the empirically found $\Delta R^2$ effect sizes; blue asterisks represent the 99th percentile $\Delta R^2$ effect sizes from MC simulations. A solid green line is placed at the predetermined threshold for practically meaningful DIF ($\Delta R^2$ = .035); a dashed blue line is placed just above the highest simulated effect size threshold.

Due to restricted variability in some items, we were unable to analyze all items. 272 out of 300 items (90.7%) in the A-IADL-Q-SV were analyzed. Of the items analyzed, 26.6% had statistically significant DIF. Effect sizes were very small for all factors ($\Delta R^2$ range .000–.034, see Figure 1). Monte Carlo simulations showed that the mean *p*-value for the chi-square statistic across all items varied between comparisons from .006 to .012, which was close to the .01 α-level used to detect DIF. Simulation-based thresholds for effect size ranged from .001 to .018 across all analyses (Figure 1). Lowering of the threshold would lead to more items being flagged for DIF. The effect sizes, however, remained very small.

For the original version, 437 out of 490 items (89.2%) were analyzed. Of those, 20.4% had statistically significant DIF. The effects for age, gender and education were again small ($\Delta R^2$ range .000–.032). Four items showed meaningful DIF for nationality with a moderate effect. In Spain, 'using the washing machine' ($\Delta R^2$ = .043), 'making appointments' ($\Delta R^2$ = .064), and 'playing card and board games' ($\Delta R^2$ = .043) were flagged. All three items had uniform DIF: the first item was more difficult for Spanish individuals, the other two were easier, as compared to the Dutch reference group. The fourth item had non-uniform DIF and was found in the French group: ‘functioning adequately at work’ (Δ$R^2$ = .064). The item appeared to be better at discriminating between people with lower and higher levels of functional impairment in France than in the Netherlands. We used the DIF results to re-estimate the *T*-scores for Spanish and French participants, thus correcting for the effect of DIF. In the Spanish group, the mean score decreased by 0.16 points on the *T*-scale, in the French group, the mean score decreased by 0.07 points. The largest individual differences in both countries (-1.14 and -1.33, respectively) corresponded to a difference of approximately one tenth of a SD, and can therefore be considered negligible. Figure 2 shows the individual score changes after DIF correction in Spain and France. There was no meaningful bias for nationality in the other countries. Simulations showed the mean chi-square statistic *p*-value across all items varied from .008 to .012. The largest Δ$R^2$ effect size was .026 (range .001–.026), which corresponds to a negligible effect.

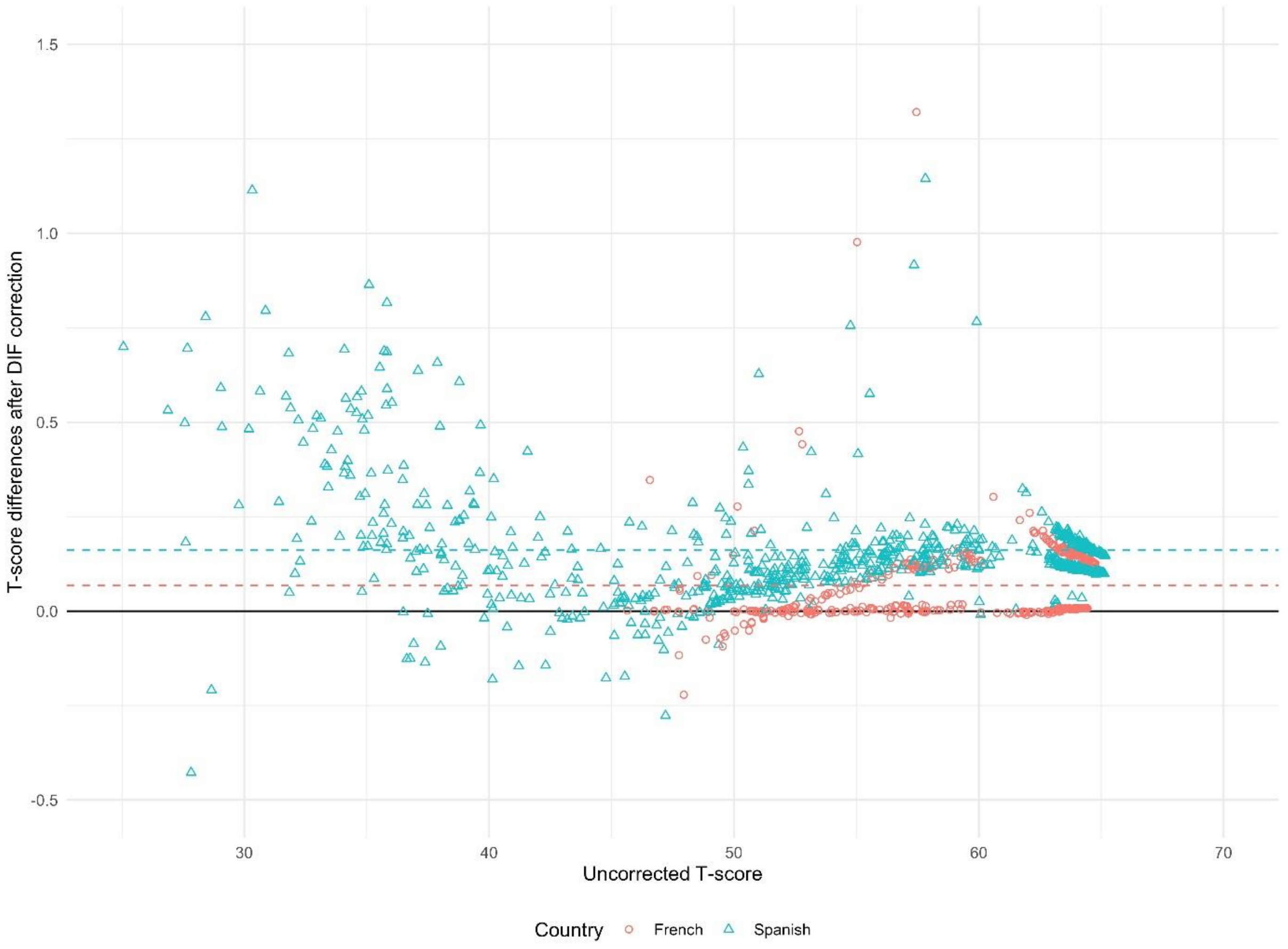

**Figure 2:** Scatter plot showing the differences between initial (uncorrected) and DIF-corrected *T*-scores for the A-IADL-Q in the French (red) and Spanish (blue) groups, plotted against the uncorrected *T*-scores.

A dashed line is placed at the mean change in score in the French and Spanish groups. Difference in total score ranges from -0.5 to +1.5 on the *T*-score, corresponding to approximately one tenth of a standard deviation difference. A solid black line is placed at no change.

### 3.3 A-IADL-Q-SV construct validation

Overall, all correlations were in the same directions and of similar magnitudes as compared to the original validation data from the Netherlands [30]. Age seemed more strongly associated with IADL impairment in Spain (*r* = -0.47, 95%CI = [-0.51, -0.42]), Greece (*r* = -0.31, 95%CI = [-0.52, -0.06]), and Serbia (*r* = -0.48, 95%CI = [-0.68, -0.21]) than in the Netherlands (*r* = -0.08, 95%CI = [-0.13, -0.02]). MMSE scores appeared to be less associated with IADL impairment France (*r* = 0.11, 95%CI = [0.02, 0.21]), United States (*r* = 0.12, 95%CI = [-0.05, 0.27]), and United Kingdom (*r* = -0.10, 95%CI = [-0.33, 0.14]), compared to the reference (*r* = 0.33, 95%CI = [0.28, 0.38]). In these countries, the MMSE had a restricted score range. Conversely, MMSE scores were more strongly associated with IADL impairment in Serbia (*r* = 0.56, 95%CI = [0.32, 0.73]). An overview of all correlations can be found in the Supplementary Material.

## 4. Discussion

In this study, we demonstrated that the influence of diversity on the measurement of IADL impairment, as measured with the A-IADL-Q, seems minimal. Although we found some differences with regard to activity endorsement between countries, there was no evidence of practically meaningful item bias caused by various factors, including age, gender, education and culture. These findings, together with the similar associations with demographic, cognitive and functional measures as found in earlier validation efforts [30], further support the validity of the A-IADL-Q.

Addressing potential bias caused by various types of diversity is highly relevant in dementia research [14]. With respect to the measurement of functional impairment, there have been contradictory findings with some studies showing a general comparability of IADLs across cultures and different ethnoracial groups [8, 9], and others reporting differences between cultures, genders, and ages [48-51]. For an optimal comparison of functional outcome in international studies and clinical trials, a valid, cross-culturally adapted instrument is crucial. In the present study, the relevance of addressing potential bias was underlined by the fact that we found some differences in activity endorsement, particularly in activities related to the household and to technology. Gender roles can differ between countries, and they might determine the IADL activities one participates in. In Mediterranean countries, it seemed people used computers less often than in Northern European countries and America.

In our current sample, the effects of DIF were small and thus did not pass our threshold for practically meaningful DIF. The reason that we found little evidence of meaningful DIF may be attributed to the cross-cultural adaptation process that all translations went through, in which potential cross-cultural differences were identified beforehand and cultural adaptations were made as necessary. These changes were minor, and we believe the items included should be applicable to Western culture in general. As part of the development of the short version, international experts provided feedback on the cross-cultural comparability of the items [33], which may explain the absence of practically meaningful item bias for nationality. Because the A-IADL-Q-SV does not appear to have practically meaningful item bias, *T*-scores do not need to be adjusted in order to be compared across countries, ages, genders, or levels of education. This suggests that the A-IADL-Q yields valid and cross-culturally comparable estimations of functional decline. Previous studies [30, 31, 33] have already shown that A-IADL-Q scores are independent of age, gender, and education, and our findings corroborate this. This is an important finding, because other functional instruments do appear to be biased for gender, age, and cultural differences [48, 49].

In the original version, a few items appeared to be biased in Spain and France. 'Making appointments' had the largest DIF effect, and a potential explanation is that examples were added in the Spanish translation, because language experts indicated that the proposed translation for the word 'appointments' (*citas*) could be interpreted as '(romantic) dates', whereas the intended definition was broader. However, adding examples may actually have restricted the interpretation of the question to the specific examples given, and led to a loss of the broader meaning. The other items with DIF had a smaller effect, and no clear reason for the presence of DIF could be discerned. Despite the finding of item bias in the original version, the effect on the total scores was minimal.

The associations between A-IADL-Q-SV scores and demographic, cognitive and functional measures we found here, largely correspond to those previously described for the original version [30]. In Spain, Greece, and Serbia, participants were older than average, and associations between age and IADL were stronger. In Spain, an association between age and IADL functioning was found earlier in a group of patients without dementia [21]. In France, the United States and the United Kingdom, the studies recruited mainly cognitively healthy participants, resulting in limited variation in the measure of cognition and IADL functioning seemed to be less associated with cognitive measures.

An important strength of this study is that we used a data-driven approach to investigate the cross-cultural comparability of IADL. We used DIF, which is a powerful procedure to detect variance in measurement between groups on an item level and was possible as a result of the IRT scoring method. Not only does DIF tell us whether an item may be biased, but it also provides insight in the impact of the bias on the overall scores and it allows for correction. We additionally used simulations to further validate the empirical findings. These advantages allowed us to create a clear picture of possible measurement variance and impact on the instrument. Another strength of the study is that we included data from more than 3,500 individuals from eight countries. People with a wide variety of cognitive impairment-related diagnoses or complaints were included, ranging from subjective cognitive decline to dementia. Furthermore, the age of participants ranged from adulthood to old age. The large sample size and large variety in diagnoses and age contributes to the generalizability of our results and conclusions.

This study also had a few limitations. First, we only included data from eight developed, Western countries. Our findings cannot be generalized to other parts of the world. One study found DIF in an IADL instrument between different Asian cultures [52]. It should also be noted that we use the term 'culture' to refer to each country's national culture. Furthermore, we did not have access to information about ethnicity or race. It is currently unclear what the influence of ethnoracial differences are on the measurement of IADL. Second, our sample was mainly comprised of highly educated people. The group we defined as having low education still received up to 12 years of education. It is possible that different results would be obtained in samples with less formal education. Third, the sample size was relatively small in Finland, Serbia, and Greece. This may have reduced our power to detect DIF. We tried to address this issue by performing Monte Carlo simulations, which indicated that the predetermined cutoff for practically meaningful DIF may have been somewhat high. More items would show DIF, if the threshold was lowered. However, when considering how these findings influence the total score, the impact seems minimal and the DIF effect sizes remain small.

The present study is an important first step in recognizing the influence of diversity on the measurement of functional impairment, and future studies should build on these findings. More research is needed to understand the differences between Western and Oriental and other cultures, as well as differences between ethnicities and races.

The A-IADL-Q-SV might be the preferred version for future international use, as it includes only the most broadly relevant everyday activities, does not seem to have meaningful item bias, has good construct validity, and is more pragmatic.

To conclude, we found no indication of the presence of clinically relevant bias caused by several aspects of diversity, including age, gender, education, and cultural differences. This is important, because it further underlines the potential of the A-IADL-Q, and the short version in particular, as an outcome measure of daily functioning in clinical practice and clinical trials.

## Acknowledgements

The Amsterdam IADL Questionnaire is free for use in all public health and not-for-profit agencies and can be obtained via https://www.alzheimercentrum.nl/professionals/amsterdam-iadl.

The development of the Amsterdam IADL Questionnaire is supported by grants from Stichting VUmc Fonds and Innovatiefonds Zorgverzekeraars. The Amsterdam Alzheimer Center is supported by Stichting Alzheimer Nederland and Stichting VUmc Fonds. The clinical database structure for the Amsterdam Dementia Cohort was developed with funding from Stichting Dioraphte. The ALFA+ Study has received funding from “la Caixa” Foundation (LCF/PR/GN17/1030004), the Alzheimer’s Association, and an international anonymous charity foundation through the TriBEKa Imaging Platform project. The work for EPAD has received support from the EU/EFPIA Innovative Medicines Initiative Joint Undertaking EPAD grant agreement n° 115736.

## Conflict of interest

The Amsterdam IADL Questionnaire was developed by SAMS and PS who were involved in the conception of the present study. The other authors report no conflict of interests.

## Highlights

- Diversity in age, gender, education and culture may influence measurement of IADL. (85)
- 3,571 people from 8 countries answered the Amsterdam IADL Questionnaire (A-IADL-Q). (85)
- Minor item bias was found for country, with a marginal influence on total scores. (81)
- No meaningful item bias was found for age, gender, and education. (67)
- These findings provide evidence for valid measurement of everyday functioning. (78)

## Supplementary Material

### Cross-cultural adaptation

To date, the Amsterdam IADL Questionnaire has been translated into thirteen languages following an extensive cross-cultural adaptation process. Some translations were made by researchers who wanted to use the questionnaire in a language that was not yet available, while others were made on request by ICON plc (https://www.iconplc.com/), a company specialized in the translation of clinical instruments.

The cross-cultural adaptation process was comprised of seven steps. First, two native speakers of the target language independently translated the questionnaire from either one of the two source languages (American English or Dutch) into the target language. Second, the two translations were reconciled into a single ‘forward translation’. Any discrepancies between the two translations were discussed and a single translation was chosen. The forward translation was subsequently translated back into the source language by two new individuals. This step was performed to check whether the intended meaning of the instructions, questions and answer options were retained. Additionally, it allowed the developers to review translations in languages they do not speak. If needed, adjustments were made in the forward translation. The fourth step was a discussion of the forward and backward translations among the translators, the developer and the translation project coordinator. This step should lead to a preliminary consensus translation (see Figure 1).

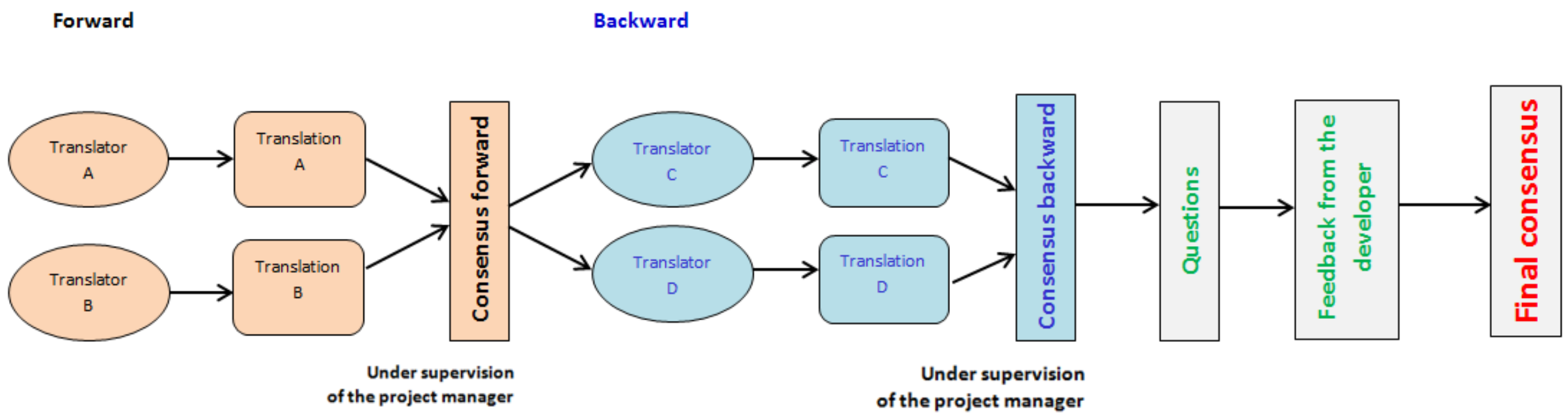

***Figure 1:*** *Translation process for the A-IADL-Q.*

Some activities included in the instrument were deemed to be less relevant for certain countries, e.g., ‘preparing sandwiches’ for Spain, or ‘using the coffee maker’ for the United Kingdom. Thus, minor changes were made to reflect the habits in the target population better: ‘preparing sandwiches’ was adapted to ‘preparing a cold meal’ in Spain and ‘or making a pot of tea’ was added to ‘using the coffee maker’ in the United Kingdom.

Subsequently, an expert committee, consisting of a small number of clinicians and knowledgeable professionals, was invited to review the translation for contents and clarity. The committee was asked to check whether the activities were clearly formulated and whether they correctly depicted the intended concepts. In the penultimate step, the translation project coordinator organized a pilot test with approximately ten caregivers to people with dementia. These caregivers should be native speakers of the target language and should not be experts in the field of questionnaires. In thinking-out-loud cognitive interviews, the caregivers were asked to explain how they would interpret and answer the questions.

In the seventh and last step, a consensus meeting was held with the translation project coordinator and the developer to discuss the feedback from the expert committee and cognitive interviews, and to address potential alterations to the translation of the items. After completion of this process, the translation was considered to be cross-culturally adapted and suitable for use. Translations performed by ICON plc were in accordance with ISO 17100:2015 regulations, and a linguistic validation certificate was also made available.

## Correlations

*Table 1*

*Pearson's* r *or Kendall's* τ *values and 95% confidence intervals for the correlations between A-IADL-Q-SV* T*-scores and demographic data and cognitive and functional measures, per country*

| Country | Age | Education | Sex | MMSE | CAMCOG | CDR | GDS |
|---|---|---|---|---|---|---|---|
| **Netherlands**[1] | -0.08 [-0.13, -0.02] | 0.09 [0.04, 0.15] | 0.04 [-0.02, 0.09] | 0.33 [0.28, 0.38] | 0.33 [0.28, 0.38] | -0.45 [-0.49, -0.41] | -0.11 [-0.16, -0.05] |
| **Spain** | -0.47 [-0.51, -0.42] | 0.34 [0.28, 0.40] | -0.01 [-0.08, 0.06] | 0.34 [0.28, 0.40] | 0.50 [0.41, 0.58] | -0.13 [-0.21, -0.05] | -0.04 [-0.12, 0.05] |
| **France** | 0.02 [-0.07, 0.10] | 0.09 [-0.01, 0.18] | -0.04 [-0.13, 0.06] | 0.11 [0.02, 0.21] | — | -0.08 [-0.17, 0.02] | -0.14 [-0.28, 0.00] |
| **United States** | 0.03 [-0.14, 0.20] | 0.07 [-0.09, 0.23] | -0.06 [-0.22, 0.09] | 0.12 [-0.05, 0.27] | — | — | -0.04 [-0.20, 0.12] |
| **United Kingdom** | -0.06 [-0.26, 0.15] | 0.01 [-0.20, 0.21] | -0.20 [-0.39, 0.00] | -0.10 [-0.33, 0.14] | — | 0.00 [-0.24, 0.23] | -0.13 [-0.33, 0.08] |
| **Greece** | -0.31 [-0.52, -0.06] | 0.06 [-0.34, 0.44] | -0.17 [-0.52, 0.24] | 0.22 [-0.18, 0.56] | 0.24 [-0.17, 0.57] | -0.44 [-0.70, -0.06] | 0.03 [-0.36, 0.41] |
| **Serbia** | -0.48 [-0.68, -0.21] | 0.10 [-0.20, 0.38] | -0.07 [-0.35, 0.23] | 0.56 [0.32, 0.73] | — | — | — |
| **Finland** | 0.01 [-0.29, 0.31] | 0.12 [-0.19, 0.41] | -0.15 [-0.43, 0.15] | 0.31 [0.01, 0.56] | — | — | -0.29 [-0.55, 0.02] |

[1] The Netherlands served as the reference group.

"—" denotes that data was not available.

*Abbreviations:* MMSE Mini-Mental State Examination; CAMCOG Cambridge Cognitive Examinations; CDR Clinical Dementia Rating; GDS Geriatric Depression Scale.